\documentclass[twocolumn,pre,superscriptaddress]{revtex4}

\usepackage{comment}
\usepackage[pdftex]{graphicx}
\usepackage{float}
\usepackage{subfigure}
\usepackage{amsmath}

\usepackage{tikz}
\usetikzlibrary{shapes.geometric, arrows}
\tikzstyle{regular} = [rectangle, rounded corners, minimum width=2cm, minimum height=1cm,text centered, draw=black, fill=white]
\tikzstyle{arrow} = [thick,->,>=stealth]
\tikzstyle{invis} = [rectangle, rounded corners, minimum width=1cm, minimum height=0cm,text centered, draw=white, fill=white]

\newcommand{\etal}{{\it{}et~al.}}
\newcommand\defn{\textit}
\newcommand\half{\tfrac12}
\newcommand{\set}[1]{\lbrace#1\rbrace}
\newcommand\mat{\mathbf}
\renewcommand\vec{\mathbf}
\newcommand\dd{\mathrm{d}}
\newcommand{\av}[1]{\left\langle#1\right\rangle}
\newcommand\Beta{\mathrm{B}}

\begin{document}

\title{Structural inference for uncertain networks}

\author{Travis Martin}
\affiliation{Department of Electrical Engineering and Computer Science,
  University of Michigan, Ann Arbor, Michigan, USA}
\author{Brian Ball}
\affiliation{Department of Physics, University of Michigan, Ann Arbor,
  Michigan, USA}
\affiliation{Quicken Loans, Detroit, Michigan, USA}
\author{M. E. J. Newman}
\affiliation{Department of Physics, University of Michigan, Ann Arbor,
  Michigan, USA}
\affiliation{Center for the Study of Complex Systems, University of
  Michigan, Ann Arbor, Michigan, USA}

\begin{abstract}
  In the study of networked systems such as biological, technological, and
  social networks the available data are often uncertain.  Rather than
  knowing the structure of a network exactly, we know the connections
  between nodes only with a certain probability.  In this paper we develop
  methods for the analysis of such uncertain data, focusing particularly on
  the problem of community detection.  We give a principled
  maximum-likelihood method for inferring community structure and
  demonstrate how the results can be used to make improved estimates of the
  true structure of the network.  Using computer-generated benchmark
  networks we demonstrate that our methods are able to reconstruct known
  communities more accurately than previous approaches based on data
  thresholding.  We also give an example application to the detection of
  communities in a protein-protein interaction network.
\end{abstract}

\maketitle

\section{Introduction}
\label{sec:intro}
Many systems of scientific interest can be usefully represented as networks
and the last few years have seen a surge of interest in the study of
networks, due in part to the fruitful application of a range of techniques
drawn from physics~\cite{Newman10}.  Most current techniques for the
analysis of networks begin with the assumption that the network data
available to us are reliable, a faithful representation of the true
structure of the network.  But many real-world data sets, perhaps most of
them, in fact contain errors and inaccuracies.  Thus, rather than
representing a network by a set of nodes joined by binary yes-or-no edges,
as is commonly done, a more realistic approach would be to specify a
probability or likelihood of connection between every pair of nodes,
representing our certainty (or uncertainty) about the existence of the
corresponding edge.  If most of the probabilities are close to zero or one
then the data are reliable---for every node pair we are close to being
certain that it either is or is not connected by an edge.  But if a
significant fraction of pairs have a probability that is neither close to
zero nor close to one then we are uncertain about the network structure.
In recent years an increasing number of network studies have started to
provide probabilistic estimates of uncertainty in this way, particularly in
the biological sciences.

One simple method for dealing with uncertain networks is
\defn{thresholding}: we assume that edges exist whenever their probability
exceeds a certain threshold that we choose.  In work on protein-protein
interaction networks, for example, Krogan~\etal~\cite{Krogan2006} assembled
a sophisticated interaction data set that includes explicit estimates of
the likelihood of interaction between every pair of proteins studied.  To
analyze their data set, however, they then converted it into a conventional
binary network by thresholding the likelihoods, followed by traditional
network analyses.  While this technique can certainly reveal useful
information, it has some drawbacks.  First, there is the issue of the
choice of the threshold level.  Krogan~\etal\ used a value of 0.273 for
their threshold, but there is little doubt that their results would be
different if they had chosen a different value and little known about how
to choose the value correctly.  Second, thresholding throws away
potentially useful information.  There is a substantial difference between
an edge with probability 0.3 and an edge with probability 0.9, but the
distinction is lost if one applies a threshold at~0.273---both fall above
the threshold and so are considered to be edges.  Third, and more subtly,
thresholded probability values fail to satisfy certain basic mathematical
requirements, meaning that thresholded networks are essentially guaranteed
to be wrong, often by a wide margin.  If, for instance, we have 100 node
pairs connected with probability~0.5 each, then on average we expect 50 of
those pairs to be connected by edges.  If we place a threshold on the
probability values at, say, 0.273, however, then all 100 of them will be
converted into edges, a result sufficiently far from the expected value of
50 as to have a very low chance of being correct.

In this paper we develop an alternative and principled approach to the
analysis of uncertain network data.  We focus in particular on the problem
of community detection in networks, one of the best studied analysis tasks.
We make use of maximum-likelihood inference techniques, whose application
to networks with definite edges is well
developed~\cite{NS01,CMN08,GZFA09,DKMZ11a}.  Here we extend those
developments to uncertain networks and show that the resulting analyses
give significantly better results in controlled tests than thresholding
methods.  As a corollary, our methods also allow us to estimate which of
the uncertain edges in a data set is mostly likely to be a true edge and
hence reconstruct, in a probabilistic fashion, the true structure of the
underlying network.

A number of authors have looked at related questions in the past.  There
exists a substantial literature on the analysis of weighted networks,
meaning networks in which the positions of the edges are exactly known but
the edges carry varying weights, such as strengths, lengths, or volumes of
traffic.  Such weighted networks are somewhat similar to the uncertain
networks studied in this paper---edges can be either strong or weak in a
certain sense---but at a deeper level they are different.  For instance,
the data sets we consider include probabilities of connection for every
node pair, whereas weighted networks have weights only for node pairs that
are known to be connected by an edge.  More importantly, in our uncertain
networks we imagine that there is a definite underlying network but that it
is not observed; all we see are noisy measurements of the underlying truth.
In weighted networks the data are considered to be exact and true and the
variation of edge weights represents an actual physical variation in the
properties of connections.

Methods for analyzing weighted networks include simple mappings to
unweighted networks and generalizations of standard methods to the weighted
case~\cite{Newman04f}.  Inference methods, akin to those we use here, have
also been applied to the weighted case~\cite{Aicher2014} and to the case of
affinity matrices, as used for example in computer vision for image
segmentation~\cite{Robles-Kelly2004}.  A little further afield, Harris and
Srinivasan~\cite{Harris14} have looked at network failures in a noisy
network model in which edges are deleted with uniform probability, while
Saade~\etal~\cite{Saade2015} use spectral techniques to detect node
properties, but not community affiliations, when the underlying network is
known but the node properties depend on noisy edge labels.  In related
work, Xu~\etal~\cite{XuML14} have studied the prediction of edge labels
using inference methods and Kurihara~\etal~\cite{kurihara2006} have applied
inference to a case where the data give the frequency of interaction
between nodes.

\section{Methods}
\label{sec:methods}
We focus on the problem of community detection in networks whose structure
is uncertain.  We suppose that we have data which, rather than specifying
with certainty whether there is an edge between two nodes $i$ and~$j$,
gives us only a likelihood or probability~$Q_{ij}$ that there is an edge.
We will assume that the probabilities are independent.  Correlated
probabilities are certainly possible, but the simple case of independent
probabilities already gives many interesting results, as we will see.

At the most basic level our goal is to classify the nodes of the network
into non-overlapping communities---groups of nodes with dense connections
within groups and sparser connections between groups, also known as
``assortative'' structure.  More generally we may also be interested in
disassortative structures in which there are more connections between
groups than within them, or mixed structures in which different groups may
be either assortative or disassortative within the same network.
Conceptually, we assume that even though our knowledge of the network is
uncertain, there is a definite underlying network in which each edge either
exists or does not, but we cannot see this network.  The underlying network
is assumed to be undirected and simple (i.e.,~it has no multi-edges or
self-edges).  The edge probabilities we observe are a noisy representation
of the true network, but they nonetheless can contain information about
structure---enough information, as we will see, to make possible the
accurate detection of communities in many situations.

Our approach to the detection problem takes the classic form of a
statistical inference algorithm.  We propose a generative model for
uncertain community-structured networks, then fit that model to our
observed data.  The parameters of the fit tell us about the community
structure.

\subsection{The model}
\label{sec:model}
The model we use is an extension to the case of uncertain networks of the
standard stochastic block model, a random graph model widely used for
community structure analyses~\cite{Holland1983,NS01,KN11a}.  In the
conventional definition of the stochastic block model, a number~$n$ of
nodes are distributed at random among~$k$ groups, with a
probability~$\gamma_r$ of being assigned to group~$r$, where $\sum_{r=1}^k
\gamma_r = 1$.  Then undirected edges are placed independently at random
between node pairs with probabilities~$\omega_{rs}$ that depend only on the
groups~$r,s$ that a pair belongs to and nothing else.  If the diagonal
elements~$\omega_{rr}$ of the probability matrix are significantly larger
than the off-diagonal entries then one has traditional assortative
community structure, with a higher density of connections within groups
than between them.  But one can also make the diagonal entries smaller to
generate disassortative structure or mixed structure types.

Given the parameters $\gamma_r$ and~$\omega_{rs}$, one can write down the
probability, or likelihood, that we generate a particular network in which
node~$i$ is assigned to group~$g_i$ and the placement of the edges is
described by an adjacency matrix~$\mat{A}$ with elements $A_{ij}=1$ if
there is an edge between nodes~$i$ and~$j$ and 0 otherwise:
\begin{align}
P(\mat{A},\vec{g}|\boldsymbol{\gamma},\boldsymbol{\omega})
 &= P(\vec{g}|\boldsymbol{\gamma}) P(\mat{A}|\vec{g},\boldsymbol{\omega})
    \nonumber\\
 &= \prod_i \gamma_{g_i} \prod_{i<j} \omega_{g_ig_j}^{A_{ij}}
    (1-\omega_{g_ig_j})^{1-A_{ij}}.
\label{eq:llA}
\end{align}
Here $\boldsymbol{\gamma}$ represents the vector of group
probabilities~$\gamma_r$ and $\boldsymbol{\omega}$ represents the matrix of
probabilities~$\omega_{rs}$.

\begin{figure*}
\centering
\begin{tikzpicture}[node distance = 4cm]
\node (model) [regular, label=Random network model]
{$P(\mat{A},\vec{g}|\boldsymbol{\gamma},\boldsymbol{\omega})$}; 
\node (network) [regular, right of=model, label=Network instance] {$\mat{A}$};
\node (noise) [invis, right of=network] {};
\node (noise e) [regular, above of=noise, label=Noise process, yshift=-3.3cm] {$\beta_1(x)$};
\node (noise ne) [regular, below of=noise, yshift=3.3cm] {$\beta_0(x)$};
\node (uncertain) [regular, right of=noise, label=Uncertain network] {$\mat{Q}$};
\draw [arrow] (model) -- (network);
\draw [arrow] (network) -- (noise e) node [midway, fill=white] {$A_{ij}=1$};
\draw [arrow] (noise e) -- (uncertain) ;
\draw [arrow] (network) -- (noise ne) node [midway, fill=white] {$A_{ij}=0$};
\draw [arrow] (noise ne) -- (uncertain);
\end{tikzpicture}
\caption{The model of uncertain network generation used in our
  calculations.  A community assignment~$\vec{g}$ and network~$\mat{A}$ are
  drawn from a random network model such as the stochastic blockmodel.  The
  experimental uncertainty is represented by giving each pair of
  nodes~$i,j$ a probability~$Q_{ij}$ of being connected by an edge, drawn
  from different distributions for edges~$A_{ij}=1$ and
  non-edges~$A_{ij}=0$.}
\label{fig:noise_flow}
\end{figure*}
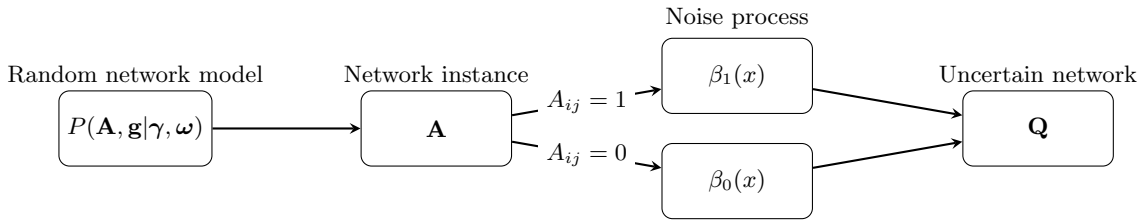

In extending the stochastic block model to uncertain networks we imagine a
multi-step process, illustrated in Fig.~\ref{fig:noise_flow}, in which the
network is first generated using the standard stochastic block model and
then the definite edges and non-edges are replaced by probabilities,
effectively adding noise to the network data.  The exact shape of the noise
will depend on the detailed effects of the experimental procedure used to
measure the network, which we assume to be unknown.  We assume only that
the edge likelihoods are true probabilities in a sense defined below (see
Eq.~\eqref{eq:noise_assumption}).  Remarkably, however, it still turns out
to be possible to perform precise inference on the data.

We represent the noise process by two unknown functions.  The function
$\beta_1(Q)$ represents the probability density on the interval from 0 to~1
that a true edge between two nodes in the original (unobserved) network
gives rise to a measured probability~$Q$ of connection between the same
nodes in the observed (probabilistic) data.  Conversely, the function
$\beta_0(Q)$ represents the probability density that a non-edge gives rise
to probability~$Q$.

Given these two functions, we can write an expression for the probability
(technically, probability density) that a true network represented by
adjacency matrix~$\mat{A}$ gives rise to a matrix of observed edge
probabilities~$\mat{Q} = \set{Q_{ij}}$ thus:
\begin{equation}
P(\mat{Q}|\mat{A}) = \prod_{i<j} \bigl[ \beta_1(Q_{ij}) \bigr]^{A_{ij}}
  \bigl[ \beta_0(Q_{ij}) \bigr]^{1-A_{ij}}.
\label{eq:PQA}
\end{equation}

The crucial observation that makes our calculations possible is that the
functions~$\beta_0$ and~$\beta_1$ are not independent, because the
numbers~$Q_{ij}$ that they generate are not just any edge weights but are
specifically probabilities and are assumed to be independent.  If we were
to gather together all node pairs that have probability~$Q$ of being
connected by an edge, the independence assumption implies that a
fraction~$Q$ of them on average should in fact be connected by edges and
the remainder should be non-edges.  For example, $90\%$ of all node pairs
with $Q_{ij}=0.9$ should, in expectation, be connected by edges.

If there are $m$ edges in total in our underlying true network, then there
are $m\beta_1(Q)\>\dd Q$ edges with observed probability lying between~$Q$
and $Q+\dd Q$ and $[{n\choose2}-m]\beta_0(Q)\>\dd Q$ non-edges in the same
interval.  Hence for every possible value of~$Q$ we must have
\begin{equation}
  \frac{m\beta_{1}(Q)\>\dd Q}{m\beta_{1}(Q)\>\dd Q
  + ({n \choose 2} - m)\beta_0(Q)\>\dd Q} = Q.
  \label{eq:noise_def}
\end{equation}
Rearranging, we then find that
\begin{equation}
\frac{\beta_1(Q)}{\beta_0(Q)} = {Q/\rho\over(1-Q)/(1-\rho)},
\label{eq:noise_assumption}
\end{equation}
where
\begin{equation}
\rho = {m\over{n\choose2}}
\label{eq:density}
\end{equation}
is the so-called \defn{density} of the network, the fraction of possible
edges that are in fact present.  Since we don't know the true network, we
don't normally know the value of~$m$, but it can be approximated by the
expected number of edges $\sum_{i<j} Q_{ij}$, which becomes an increasingly
good estimate as the network gets larger, and from this figure we can
calculate~$\rho$.

Note that Eq.~\eqref{eq:noise_assumption} implies that $\beta_0(1)=0$ and
$\beta_1(0)=0$.  The equation is also compatible with the choice
$\beta_0(Q)=\delta(Q)$, $\beta_1(Q)=\delta(Q-1)$, where $\delta(x)$ is the
Dirac delta function, which corresponds to the conventional case of a
perfectly certain network with $Q_{ij} = A_{ij}$.

Using Eq.~\eqref{eq:noise_assumption} we can now write Eq.~\eqref{eq:PQA}
as
\begin{align}
P(\mat{Q}|\mat{A}) &= \prod_{i<j} {1-\rho\over1-Q_{ij}}\beta_0(Q_{ij})
   \nonumber\\
  &\qquad\times \prod_{i<j} \biggl( {Q_{ij}\over\rho} \biggr)^{A_{ij}}
  \biggl( {1-Q_{ij}\over1-\rho} \biggr)^{1-A_{ij}}.
\label{eq:PQA2}
\end{align}
The first product is a constant for any given set of observed
probabilities~$\mat{Q}$ and hence will have no effect on our
maximum-likelihood calculations (which depend only on the position of the
likelihood maximum and not on its absolute value).  Henceforth, we will
neglect this factor.  Then we combine Eqs.~\eqref{eq:llA}
and~\eqref{eq:PQA2} to get an expression for the likelihood of the
data~$\mat{Q}$ and the community assignments~$\vec{g}$, neglecting
constants and given the model parameters~$\boldsymbol{\gamma}$
and~$\boldsymbol{\omega}$:
\begin{align}
P(\mat{Q},\vec{g}|\boldsymbol{\gamma},\boldsymbol{\omega})
  &= \sum_{\mat{A}} P(\mat{Q}|\mat{A})
     P(\mat{A},\vec{g}|\boldsymbol{\gamma},\boldsymbol{\omega}) \nonumber\\
  &\hspace{-5.5em}{} = \prod_i \gamma_{g_i}
     \prod_{i<j} \sum_{A_{ij}=0,1}
     \biggl[ {Q_{ij}\omega_{g_ig_j}\over\rho} \biggr]^{A_{ij}}
     \biggl[ {(1-Q_{ij})(1-\omega_{g_ig_j})\over1-\rho} \biggr]^{1-A_{ij}}
     \nonumber\\
  &\hspace{-5.5em}{} = \prod_i \gamma_{g_i}
   \prod_{i<j}
     \biggl[ {Q_{ij}\omega_{g_ig_j}\over\rho} +
     {(1-Q_{ij})(1-\omega_{g_ig_j})\over1-\rho} \biggr].
\label{eq:pqggw}
\end{align}

Our goal is now, given a particular set of observed data~$\mat{Q}$, to
maximize this likelihood to find the best-fit
parameters~$\boldsymbol{\gamma}$ and~$\boldsymbol{\omega}$.  In the process
we will determine the community assignments~$\vec{g}$ as well (which are
frequently the primary objects of interest).

\subsection{Fitting to empirical data}
\label{sec:analysis}
Fitting the model to an observed but uncertain network, represented by the
probabilities~$Q_{ij}$, means determining the values of the
parameters~$\boldsymbol{\gamma}$ and~$\boldsymbol{\omega}$ that maximize
the probability of generating the particular data we see.  In other words,
we want to maximize the \defn{marginal likelihood} of the data given the
parameters:
\begin{equation}
P(\mat{Q}|\boldsymbol{\gamma},\boldsymbol{\omega})
  = \sum_{\vec{g}}
    P(\mat{Q},\vec{g}|\boldsymbol{\gamma},\boldsymbol{\omega}).
\label{eq:llmarginal}
\end{equation}
Equivalently, we can maximize the logarithm of this quantity, which gives
the same result (since the logarithm is a monotone function) but is often
easier.

Direct maximization by differentiation gives rise to a set of implicit
equations that have no simple solution, so instead we employ a standard
trick from the statistics toolbox and apply \defn{Jensen's inequality},
which says that for any set of positive-definite quantities~$x_i$, the log
of their sum satisfies
\begin{equation}
\log \sum_i x_i \ge \sum_i q_i \log {x_i\over q_i},
\label{eq:jensen}
\end{equation}
where $q_i$ is any probability distribution over~$i$ satisfying the
normalization condition $\sum_i q_i = 1$.  One can easily verify that the
exact equality is achieved by choosing
\begin{equation}
q_i = {x_i\over\sum_i x_i}.
\label{eq:equality}
\end{equation}

\begin{widetext}
Applying Jensen's inequality to~\eqref{eq:llmarginal}, we get
\begin{align}
\log P(\mat{Q}|\boldsymbol{\gamma},\boldsymbol{\omega})
  &\ge \sum_{\vec{g}} q(\vec{g})
      \log {P(\mat{Q},\vec{g}|\boldsymbol{\gamma},\boldsymbol{\omega})\over
            q(\vec{g})} \nonumber\\
  &= \sum_{\vec{g}} q(\vec{g}) \sum_i \log \gamma_{g_i}
     + \half\sum_{\vec{g}} q(\vec{g}) \sum_{ij}
     \log \biggl[ {Q_{ij}\omega_{g_ig_j}\over\rho} +
     {(1-Q_{ij})(1-\omega_{g_ig_j})\over1-\rho} \biggr]
     - \sum_{\vec{g}} q(\vec{g}) \log q(\vec{g}) \nonumber\\
  &= \sum_i \sum_r q_r^i \log \gamma_r
     + \half \sum_{ij} \sum_{rs} q_{rs}^{ij}
     \log \biggl[ {Q_{ij}\omega_{rs}\over\rho} +
     {(1-Q_{ij})(1-\omega_{rs})\over1-\rho} \biggr]
     - \sum_{\vec{g}} q(\vec{g}) \log q(\vec{g}),
\label{eq:inequality}
\end{align}
\end{widetext}
where $q_r^i$ is the marginal probability within the probability
distribution~$q(\vec{g})$ that node~$i$ belongs to community~$r$:
\begin{equation}
q_r^i = \sum_{\vec{g}} q(\vec{g}) \delta_{g_i,r},
\label{eq:onevertex}
\end{equation}
and $q_{rs}^{ij}$ is the joint marginal probability that nodes~$i$ and~$j$
belong to communities~$r$ and~$s$ respectively:
\begin{equation}
q_{rs}^{ij} = \sum_{\vec{g}} q(\vec{g}) \delta_{g_i,r} \delta_{g_j,s},
\label{eq:twovertex}
\end{equation}
with $\delta_{ij}$ being the Kronecker delta.

Following Eq.~\eqref{eq:equality}, the exact equality
in~\eqref{eq:inequality}, and hence the maximum of the right-hand side, is
achieved when
\begin{align}
q(\vec{g}) &= {P(\mat{Q},\vec{g}|\boldsymbol{\gamma},\boldsymbol{\omega})
  \over \sum_{\vec{g}}
  P(\mat{Q},\vec{g}|\boldsymbol{\gamma},\boldsymbol{\omega})} \nonumber\\
  &={\prod_i \gamma_{g_i} \prod_{i<j}
     \Bigl[ {Q_{ij}\omega_{g_ig_j}\over\rho} +
     {(1-Q_{ij})(1-\omega_{g_ig_j})\over1-\rho} \Bigr]\over
  \sum_{\vec{g}} \prod_i \gamma_{g_i} \prod_{i<j}
     \Bigl[ {Q_{ij}\omega_{g_ig_j}\over\rho} +
     {(1-Q_{ij})(1-\omega_{g_ig_j})\over1-\rho} \Bigr]}.
\label{eq:estep}
\end{align}
Thus, calculating the maximum of the left-hand side
of~\eqref{eq:inequality} with respect to the
parameters~$\boldsymbol{\gamma},\boldsymbol{\omega}$ is equivalent to a
double maximization of the right-hand side with respect to~$q(\vec{g})$ (by
choosing the value above) so as to make the two sides equal, and then with
respect to the parameters.  At first sight, this seems to make the problem
more complex, but numerically it is in fact easier---the double
maximization can be achieved in a relatively straightforward manner by
alternately maximizing with respect to~$q(\vec{g})$ using
Eq.~\eqref{eq:estep} and then with respect to the parameters.  Such
alternate maximizations can trivially be shown always to converge to a
local maximum of the log-likelihood.  They are not guaranteed to find the
global maximum, however, so commonly we repeat the entire calculation
several times from different starting points and choose among the results
the one which gives the highest value of the likelihood.

Once we have converged to the maximum, the final value of the probability
distribution~$q(\vec{g})$ is given by Eq.~\eqref{eq:estep} to be
\begin{equation}
q(\vec{g}) = {P(\mat{Q},\vec{g}|\boldsymbol{\gamma},\boldsymbol{\omega})
  \over P(\mat{Q}|\boldsymbol{\gamma},\boldsymbol{\omega})}
  = P(\vec{g}|\mat{Q},\boldsymbol{\gamma},\boldsymbol{\omega}).
\end{equation}
In other words, $q(\vec{g})$~is the posterior distribution over community
assignments~$\vec{g}$ given the observed data~$\mat{Q}$ and the model
parameters.  Thus, in addition to telling us the values of the parameters,
our calculation tells us the probability of any assignment of nodes to
communities.  Specifically, the one-node marginal probability~$q_r^i$,
Eq.~\eqref{eq:onevertex}, tells us the probability that node~$i$ belongs to
community~$r$ and, armed with this information, we can calculate the most
probable community that each node belongs to, which is the primary goal of
our calculation.

We still need to perform the maximization of~\eqref{eq:inequality} over the
parameters.  We note first that the final sum is independent of
either~$\boldsymbol{\gamma}$ or~$\boldsymbol{\omega}$ and hence can be
neglected.  Maximization of the remaining terms with respect
to~$\boldsymbol{\gamma}$ is straightforward.  Differentiating with respect
to~$\gamma_r$, subject to the normalization condition~$\sum_r \gamma_r =
1$, gives
\begin{equation}
\gamma_r = {1\over n} \sum_i q_r^i.
\label{eq:gamma}
\end{equation}

\begin{widetext}
  Maximization with respect to~$\boldsymbol{\omega}$ is a little more
  tricky.  Only the second term in~\eqref{eq:inequality} depends
  on~$\boldsymbol{\omega}$, but direct differentiation of this term yields
  a difficult equation, so instead we apply Jensen's
  inequality~\eqref{eq:jensen} again, giving
\begin{equation}
\sum_{ij} \sum_{rs} q_{rs}^{ij} \log \biggl[ {Q_{ij}\omega_{rs}\over\rho}
  + {(1-Q_{ij})(1-\omega_{rs})\over1-\rho} \biggr]
\ge \sum_{ij} \sum_{rs} q_{rs}^{ij}
  \biggl[ t_{rs}^{ij} \log {Q_{ij}\omega_{rs}\over\rho t_{rs}^{ij}}
  + (1-t_{rs}^{ij})
  \log {(1-Q_{ij})(1-\omega_{rs})\over(1-\rho)(1-t_{rs}^{ij})} \biggr],
\label{eq:ineq2}
\end{equation}
where $t_{rs}^{ij}$ is any number between zero and one.
\end{widetext}
The exact equality, and hence the maximum of the right-hand side, is
achieved when
\begin{equation}
t_{rs}^{ij} =
  {Q_{ij}\omega_{rs}/\rho\over
   Q_{ij}\omega_{rs}/\rho + (1-Q_{ij})(1-\omega_{rs})/(1-\rho)}.
\label{eq:estep2}
\end{equation}
Thus, by the same argument as previously, we can maximize the left-hand
side of~\eqref{eq:ineq2} by repeatedly maximizing the right-hand side with
respect to~$t_{rs}^{ij}$ using Eq.~\eqref{eq:estep2} and with respect
to~$\omega_{rs}$ by differentiation.  Performing the derivative and setting
the result to zero, we find that the maximum with respect to~$\omega_{rs}$
falls~at
\begin{equation}
\omega_{rs} = {\sum_{ij} q_{rs}^{ij} t_{rs}^{ij}\over\sum_{ij} q_{rs}^{ij}}.
\label{eq:mstep2}
\end{equation}
The optimal values of the~$\omega_{rs}$ can now be calculated by iterating
Eqs.~\eqref{eq:estep2} and~\eqref{eq:mstep2} alternately to convergence
from a suitable initial condition.

The quantity~$t_{rs}^{ij}$ has a simple physical interpretation, as we can
see by applying Eq.~\eqref{eq:noise_assumption} to~\eqref{eq:estep2},
giving
\begin{equation}
t_{rs}^{ij} = {\omega_{rs} \beta_1(Q_{ij})\over
  \omega_{rs} \beta_1(Q_{ij}) + (1-\omega_{rs}) \beta_0(Q_{ij})}.
\end{equation}
But by definition
\begin{align}
\omega_{rs} &= P(A_{ij}=1|g_i=r,g_j=s), \\
\beta_1(Q_{ij}) &= P(Q_{ij}|A_{ij}=1), \\
\beta_0(Q_{ij}) &= P(Q_{ij}|A_{ij}=0),
\end{align}
and hence
\begin{align}
t_{rs}^{ij} &= {P(A_{ij}=1|g_i=r,g_j=s) P(Q_{ij}|A_{ij}=1)\over
  P(Q_{ij}|g_i=r,g_j=s)} \nonumber\\
  &= P(A_{ij}=1|Q_{ij},g_i=r,g_j=s).
\label{eq:tinterp}
\end{align}
In other words, $t_{rs}^{ij}$ is the posterior probability that there is an
edge between nodes $i$ and~$j$, given that they are in groups $r$ and $s$
respectively.  This quantity will be useful shortly when we consider the
problem of reconstructing a network from uncertain observations.

We now have a complete algorithm for fitting our model to the observed
data.  The steps of the algorithm are as follows:
\begin{enumerate}
\item Make an initial guess (for instance at random) for the values of the
  parameters~$\boldsymbol{\gamma}$ and~$\boldsymbol{\omega}$.
\item Calculate the distribution~$q(\vec{g})$ from Eq.~\eqref{eq:estep}.
\item Calculate the one- and two-node marginal probabilities~$q_r^i$
  and~$q_{rs}^{ij}$ from Eqs.~\eqref{eq:onevertex}
  and~\eqref{eq:twovertex}.
\item From these quantities calculate updated values
  of~$\boldsymbol{\gamma}$ from Eq.~\eqref{eq:gamma} and
  $\boldsymbol{\omega}$ by iterating Eqs.~\eqref{eq:estep2}
  and~\eqref{eq:mstep2} to convergence starting from the current estimate
  of~$\boldsymbol{\omega}$.
\item Repeat from step~2 until $q(\vec{g})$ and the model parameters
  converge.
\end{enumerate}
Algorithms of this type are known as expectation--maximization or EM
algorithms~\cite{DLR77,MK08}.  The end result is a maximum likelihood
estimate of the parameters~$\boldsymbol{\gamma}$ and~$\boldsymbol{\omega}$
along with the posterior distribution over community
assignments~$q(\vec{g})$ and the probability~$t_{rs}^{ij}$ of an edge
between any pair of nodes.

Equation~\eqref{eq:mstep2} can usefully be simplified a little further, in
two ways.  First, note that Eq.~\eqref{eq:estep2} implies that
$t_{rs}^{ij}=0$ whenever $Q_{ij}=0$.  All of the real-world data sets we
have examined are \defn{sparse}, meaning that a large majority of the
probabilities~$Q_{ij}$ are zero.  This means that most of the terms in the
numerator of~\eqref{eq:mstep2} vanish and can be dropped from the sum,
which speeds up the calculation considerably.  Indeed $t_{rs}^{ij}$ need
not be evaluated at all for node pairs~$i,j$ such that~$Q_{ij}=0$, since
this sum is the only place that $t_{rs}^{ij}$ appears in our calculation.
Moreover it turns out that we need not evaluate $q_{rs}^{ij}$ for such node
pairs either.  The only other place that $q_{rs}^{ij}$ appears is in the
denominator of Eq.~\eqref{eq:mstep2}, which can be simplified by using
Eq.~\eqref{eq:twovertex} to rewrite it thus:
\begin{equation}
\sum_{ij} q_{rs}^{ij}
  = \sum_g q(\vec{g}) \sum_i \delta_{g_i,r} \sum_j \delta_{g_j,s}
  = \av{n_r n_s},
\end{equation}
where $\av{\ldots}$ indicates an average over~$q(\vec{g})$ and $n_r=\sum_i
\delta_{g_i,r}$ is the number of nodes in group~$r$, for community
assignment~$\vec{g}$.  For large networks the number of nodes in a group
becomes tightly peaked about its mean value so that $\av{n_r
  n_s}\simeq\av{n_r}\av{n_s}$ where $\av{n_r} = \sum_{\vec{g}} q(\vec{g})
\sum_i \delta_{g_i,r} = \sum_i q_r^i$.  Hence
\begin{equation}
\omega_{rs} = {\sum_{ij} q_{rs}^{ij} t_{rs}^{ij}\over
               \sum_i q_r^i \sum_j q_s^j}.
\label{eq:omegars}
\end{equation}
This obviates the need to calculate~$q_{rs}^{ij}$ for node pairs such
that~$Q_{ij}=0$ (which is most node pairs), and in addition speeds the
calculation further because the denominator can now be evaluated in time
proportional to the number of nodes in the network, rather than the number
of nodes squared, as in Eq.~\eqref{eq:mstep2}.  (And the numerator can be
evaluated in time proportional to the number of nonzero~$Q_{ij}$, which is
small.)

\subsection{Belief propagation}
\label{sec:bp}
In principle, the methods of the previous section constitute a complete
algorithm for fitting our model to observed network data.  In practice,
however, it is an impractical one because it's unreasonably slow.  The
bottle\-neck is the sum in the denominator of Eq.~\eqref{eq:estep}, which
is a sum over all possible assignments~$\vec{g}$ of nodes to communities.
If there are $n$ nodes and $k$ communities then there are $k^n$ possible
assignments, a number that grows with~$n$ so rapidly as to prohibit
explicit numerical evaluation of the sum for all but the smallest of
networks.

This is not a new problem.  It is common to most EM algorithms, not only
for network applications but for statistics in general.  The traditional
way around it is to approximate the distribution~$q(\vec{g})$ by importance
sampling using Markov chain Monte Carlo.  In this paper, however, we use a
different method, proposed recently by Decelle~\etal~\cite{DKMZ11a,DKMZ11b}
and specific to networks, namely belief propagation.

Originally developed in physics and computer science for the probabilistic
solution of problems on graphs and lattices~\cite{Pearl88,MM09}, belief
propagation is a message passing method in which the nodes of a network
exchange messages or ``beliefs,'' which are probabilities representing the
current best estimate of the solution to the problem of interest.  In the
present case we define a message~$\eta_r^{i\to j}$ which is equal to the
probability that node~$i$ belongs to community~$r$ if node~$j$ is removed
from the network.  The removal of a node is crucial, since it allows us to
write a self-consistent set of equations satisfied by the messages, whose
solution gives us the distribution~$q(\vec{g})$ over group assignments.
Although the equations can without difficulty be written exactly and in
full, we will here approximate them to leading order only in the small
quantities~$\omega_{rs}$.  We find this approximation to give excellent
results in our applications and the equations are considerably simpler, as
well as giving a faster final algorithm.

\begin{widetext}
  Within this approximation, the belief propagation equation for the
  message~$\eta_r^{i\to j}$ is:
\begin{equation}
\eta_r^{i\to j} = {\gamma_r\over Z_{i\to j}}
   \exp \biggl( - \sum_{k,s} q_s^k \omega_{rs} \biggr)
   \prod_{\substack{k(\ne j)\\ Q_{ik}\ne 0}}
   \sum_s \eta_s^{k\to i} \biggl[ {Q_{ik} \omega_{rs}\over\rho}
   + {(1-Q_{ik})(1-\omega_{rs})\over1-\rho} \biggr],
\label{eq:bp}
\end{equation}
where $Z_{i\to j}$ is a normalization coefficient that ensures~$\sum_r
\eta_r^{i\to j} = 1$, having value
\begin{equation}
Z_{i\to j} = \sum_r \gamma_r
   \exp \biggl( - \sum_{k,s} q_s^k \omega_{rs} \biggr)
   \prod_{\substack{k(\ne j)\\ Q_{ik}\ne 0}}
   \sum_s \eta_s^{k\to i} \biggl[ {Q_{ik} \omega_{rs}\over\rho}
   + {(1-Q_{ik})(1-\omega_{rs})\over1-\rho} \biggr],
\label{eq:Zij}
\end{equation}
and $q_r^i$ is, as before, the one-node marginal probability of
Eq.~\eqref{eq:onevertex}, which can itself be conveniently calculated
directly from the messages~$\eta_r^{i\to j}$ via
\begin{equation}
q_r^i = {\gamma_r\over Z_i}
   \exp \biggl( - \sum_{j,s} q_s^j \omega_{rs} \biggr)
   \prod_{\substack{j\\ Q_{ij}\ne 0}}
   \sum_s \eta_s^{j\to i} \biggl[ {Q_{ij} \omega_{rs}\over\rho}
   + {(1-Q_{ij})(1-\omega_{rs})\over1-\rho} \biggr],
\label{eq:bp1vertex}
\end{equation}
with
\begin{equation}
Z_i = \sum_r \gamma_r
   \exp \biggl( - \sum_{j,s} q_s^j \omega_{rs} \biggr)
   \prod_{\substack{j\\ Q_{ij}\ne 0}}
   \sum_s \eta_s^{j\to i} \biggl[ {Q_{ij} \omega_{rs}\over\rho}
   + {(1-Q_{ij})(1-\omega_{rs})\over1-\rho} \biggr].
\label{eq:Zi}
\end{equation}
\end{widetext}

These equations are exact if the set of node pairs~$i,j$ with edge
probabilities $Q_{ij}>0$ forms a tree or is at least locally tree-like
(meaning that arbitrarily large local neighborhoods take the form of trees
in the limit of large network size).  For non-trees, which includes most
real-world networks, they are only approximate, but previous results from a
number of studies show the approximation to be a good one in
practice~\cite{MM09,DKMZ11a,DKMZ11b,Yan14,ZMN15}.

Solution of these equations is by iteration.  Typically we start from the
current best estimate of the values of the beliefs and iterate to
convergence, then from the converged values we calculate the crucial
two-node marginal probability~$q_{rs}^{ij}$ by noting that
\begin{align}
q_{rs}^{ij} &= P(g_i=r,g_j=s|Q_{ij}) \nonumber\\
            &= {P(g_i=r,g_j=s) P(Q_{ij}|g_i=r,g_j=s)\over
                \sum_{rs} P(g_i=r,g_j=s) P(Q_{ij}|g_i=r,g_j=s)}.
\label{eq:qrsij1}
\end{align}
where all data~$\mat{Q}$ other than~$Q_{ij}$ are assumed given in each
probability.  The probabilities in these expressions are equal to
\begin{align}
P(g_i=r,g_j=s) &= \eta_r^{i\to j} \eta_s^{j\to i}, \\
P(Q_{ij}|g_i=r,g_j=s) &= \beta_0(Q_{ij}) {1-\rho\over1-Q_{ij}} \nonumber\\
  &\hspace{-3em}{}\times \biggl[ {Q_{ij} \omega_{rs}\over\rho}
  + {(1-Q_{ij})(1-\omega_{rs})\over1-\rho} \biggr].
\end{align}
Substituting these into~\eqref{eq:qrsij1}, we get
\begin{equation}
q_{rs}^{ij} = {\eta_r^{i\to j} \eta_s^{j\to i}
               \Bigl[ {Q_{ij} \omega_{rs}\over\rho}
                + {(1-Q_{ij})(1-\omega_{rs})\over1-\rho} \Bigr]\over
               \sum_{rs} \eta_r^{i\to j} \eta_s^{j\to i}
               \Bigl[ {Q_{ij} \omega_{rs}\over\rho}
                + {(1-Q_{ij})(1-\omega_{rs})\over1-\rho} \Bigr]}.
\label{eq:qrsij2}
\end{equation}

Our final algorithm then consists of alternately (a)~iterating the belief
propagation equations~\eqref{eq:bp} to convergence and using the results to
calculate the marginal probabilities~$q_r^i$ and~$q_{rs}^{ij}$ from
Eqs.~\eqref{eq:bp1vertex} and~\eqref{eq:qrsij2}, and (b)~iterating
Eqs.~\eqref{eq:estep2} and~\eqref{eq:omegars} to convergence to calculate
new values of the~$\omega_{rs}$ and using Eq.~\eqref{eq:gamma} to calculate
new values of~$\gamma_r$.  In practice the algorithm is efficient---in
other tests of belief propagation it has been found fast enough for
applications to networks of a million nodes or more.

\subsection{Degree-corrected model}
\label{sec:dcsbm}
Our method gives a complete algorithm for fitting the standard stochastic
block model to uncertain network data represented by the matrix~$\mat{Q}$
of edge probabilities.  As pointed out previously by Karrer and
Newman~\cite{KN11a}, however, the stochastic block model gives poor
performance for community detection on many real-world networks because the
model assumes a Poisson degree distribution, which is strongly in conflict
with the broad, frequently fat-tailed degree distributions seen in
real-world networks.  Because of this conflict it is often not possible to
find a good fit of the stochastic block model to observed network data, for
any parameter values, and in such cases the model can return poor
performance on community detection tasks.

The fix for this problem is straightforward.  The \defn{degree-corrected
  stochastic block model} is identical to the standard block model except
that the probability of an edge between nodes~$i,j$ that fall in
groups~$r,s$ is $d_i d_j \omega_{rs}$ (instead of just~$\omega_{rs}$),
where $d_i$ is the observed degree of node~$i$ in the network.  This
modification allows the model to accurately fit arbitrary degree
distributions, and community detection algorithms that perform fits to the
degree-corrected model are found to return excellent results in real-world
applications~\cite{KN11a}.

We can make the same modification to our methods as well.  The developments
follow exactly the same lines as for the ordinary (uncorrected) stochastic
block model.  The crucial equations~\eqref{eq:estep2}
and~\eqref{eq:omegars} become
\begin{equation}
t_{rs}^{ij} =
  {Q_{ij} d_i d_j \omega_{rs}/\rho\over
   Q_{ij} d_i d_j \omega_{rs}/\rho
   + (1-Q_{ij})(1-d_i d_j \omega_{rs})/(1-\rho)}
\end{equation}
and
\begin{equation}
\omega_{rs} = {\sum_{ij} q_{rs}^{ij} t_{rs}^{ij}\over
               \sum_i d_i q_r^i \sum_j d_j q_s^j},
\end{equation}
while the belief propagation equation~\eqref{eq:bp} becomes
\begin{align}
&\eta_r^{i\to j} = {\gamma_r\over Z_{i\to j}}
   \exp \biggl( - d_i d_j \sum_{k,s} q_s^k \omega_{rs} \biggr) \nonumber\\
   &\times \prod_{\substack{k(\ne j)\\ Q_{ik}\ne 0}}
   \sum_s \eta_s^{k\to i} \biggl[ {Q_{ik}d_id_j\omega_{rs}\over\rho}
   + {(1-Q_{ik})(1-d_id_j\omega_{rs})\over1-\rho} \biggr],
\end{align}
with corresponding modifications to Eqs.~\eqref{eq:Zij} to~\eqref{eq:Zi}
and Eq.~\eqref{eq:qrsij2}.

In the following sections we describe a number of example applications of
our methods.  Among these, the tests on synthetic networks
(Section~\ref{sec:synthetic}) are performed using the standard stochastic
block model, without degree-correction, while the tests on real-world
networks (Section~\ref{sec:ppi}) use the degree-corrected version.

\section{Results}
We have tested the methods described in the previous sections both on
computer-generated benchmark networks with known structure and on
real-world examples.

\subsection{Synthetic networks}
\label{sec:synthetic}
Computer-generated or ``synthetic'' networks provide a controlled test of
the performance of our algorithm.  We generate networks with known
community structure planted within them and then test whether the algorithm
is able accurately to detect that structure.

For the tests reported here, we generate networks using the standard (not
degree-corrected) stochastic block model and then add noise to them to
represent the network uncertainty, using functions~$\beta_0$ and~$\beta_1$
as defined in Section~\ref{sec:model}.  We use networks of size $n=4000$
nodes, divided into two equally-size communities, and as the noise
function~$\beta_1(Q)$ for the edges we use a beta distribution:
\begin{equation}
\beta_1(Q) = {Q^{a_1-1} (1-Q)^{b_1-1}\over\Beta(a_1,b_1)},
\label{eq:beta1}
\end{equation}
where $\Beta(a,b)$ is Euler's beta function.  As the noise
function~$\beta_0(Q)$ for the non-edges we use a beta function plus an
additional delta-function spike at zero:
\begin{equation}
\beta_0(Q) = c {Q^{a_0-1} (1-Q)^{b_0-1}\over\Beta(a_0,b_0)} + (1-c)\delta(Q).
\label{eq:beta0}
\end{equation}
The delta function makes the matrix~$\mat{Q}$ of edge probabilities
realistically sparse, in keeping with the structure of real-world data
sets, with a fraction~$1-c$ of non-edges having exactly zero probability in
the observed data, on average.

Thus there are a total of five parameters in our noise functions: $a_0$,
$b_0$, $a_1$, $b_1$, and~$c$.  Not all of these parameters are independent,
however, because our functions still have to satisfy the
constraint~\eqref{eq:noise_assumption}.  Substituting Eqs.~\eqref{eq:beta1}
and~\eqref{eq:beta0} into~\eqref{eq:noise_assumption}, we see that for the
constraint to be satisfied for all~$Q>0$ we must have $a_0=a_1-1$,
$b_0=b_1+1$, and
\begin{align}
c &= {1-\rho\over\rho}\,{\Beta(a_1,b_1)\over\Beta(a_0,b_0)}
   = {1-\rho\over\rho}\,{\Beta(a_1,b_1)\over\Beta(a_1-1,b_1+1)} \nonumber\\
  &= {1-\rho\over\rho}\,{a_1-1\over b_1}.
\end{align}
Thus there are really just two degrees of freedom in the choice of the
noise functions.  Once we fix the parameters $a_1$ and~$b_1$, everything
else is fixed also.  Alternatively, we can fix the parameter~$c$, thereby
fixing the density of the data matrix~$\mat{Q}$, plus one or other of the
parameters $a_1$ and~$b_1$.

The networks we generate are now analyzed using the non-degree-corrected
algorithm of Sections~\ref{sec:model} to~\ref{sec:bp}.  To quantify
performance we assign each node~$i$ to the community~$r$ for which its
probability~$q_r^i$ of membership, Eq.~\eqref{eq:onevertex}, as computed by
the algorithm, is greatest, then compare the result to the known true
community assignments from which the network was generated.  Success (or
lack of it) is quantified by computing the fraction of nodes placed by the
algorithm in the correct groups.  We also compare the results against the
naive (but common) thresholding method discussed in the
introduction~\cite{Krogan2006}, in which edge probabilities~$Q_{ij}$ are
turned into binary yes-or-no edges by cutting them off at some fixed
threshold~$\tau$, so that the adjacency matrix element~$A_{ij}$ is $1$ if
and only if $Q_{ij}>\tau$.  Community structure in the thresholded network
is analyzed using the standard stochastic block model algorithm described
in, for example, Refs.~\cite{DKMZ11a} and~\cite{DKMZ11b}.

\begin{figure}
\centering
\includegraphics[width=\columnwidth]{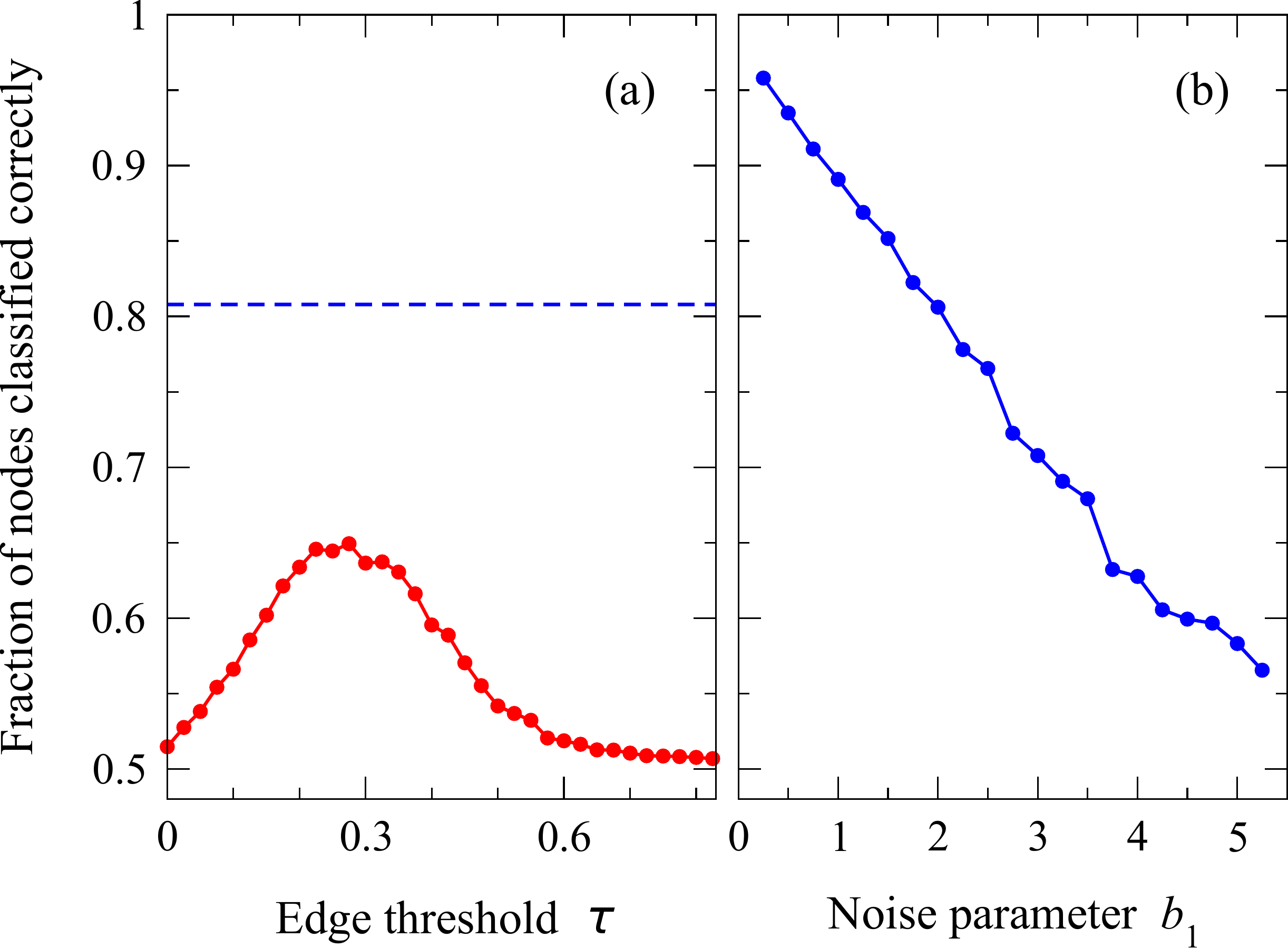}
\caption{Tests of the method described in this paper on synthetic benchmark
  networks.  (a)~Fraction of nodes placed in the correct community for
  uncertain networks generated using a stochastic block model with $n=4000$
  nodes, two groups of equal size, edge probabilities
  $\omega_{11}=\omega_{22}=0.02$, $\omega_{12}=\omega_{21}=0.014$, and
  noise parameters $a_1=1.4$ and $b_1=2$ (see Eq.~\eqref{eq:beta1}).  The
  horizontal dashed line shows the performance of the algorithm described
  in this paper.  The points show the performance of a naive algorithm in
  which the uncertain network is first converted to a binary network by
  thresholding the edge probabilities and the result then fed into a
  standard community detection algorithm.  The results for each algorithm
  are averaged over 20 repetitions of the experiment with different
  networks.  Statistical errors are comparable in size to the data points.
  (b)~Fraction of nodes classified into their correct communities for
  stochastic block model networks with varying amounts of noise in the
  data.  The parameters are the same as for (a) but with the sparsity
  parameter~$c$ fixed at $1/4n$ (see Eq.~\eqref{eq:beta0} and the ensuing
  discussion) and varying the parameter~$b_1$, which controls the level of
  noise in the data.}
\label{fig:sbm}
\end{figure}

As we vary the parameters of the underlying network and noise functions the
performance of both algorithms varies.  When the community structure is
strong and the noise is weak both algorithms (not surprisingly) do well,
recovering the community structure nearly perfectly, while for weak enough
community structure or strong noise neither algorithm does better than
chance.  But, as shown in Fig.~\ref{fig:sbm}a, there is a regime of
intermediate structure and noise in which our algorithm does significantly
better than the naive technique.  The figure shows the fraction of
correctly classified nodes in the naive algorithm as a function of the
threshold~$\tau$ (data points in the figure) compared against the
performance of the algorithm of this paper (dashed line) and, as we can
see, the latter outperforms the former no matter what value of~$\tau$ is
used.  Note that the worst possible performance still classifies a half of
the nodes correctly---even a random coin toss would get this many
right---so this is the minimum value on the plot.  For high threshold
values~$\tau$ approaching one, the threshold method throws away essentially
all edges, leaving itself no data to work with, and hence does little
better than chance.  Conversely for low thresholds the threshold method
treats any node pair with a nonzero connection probability~$Q_{ij}$ as
having an edge, even when an edge is wildly unlikely, thereby introducing
large amounts of noise into the calculation that again reduce performance
to a level little better than chance.  The optimal performance falls
somewhere between these two extremes, around $\tau=0.25$ in this case, but
even at this optimal point the thresholding method's performance falls far
short of the algorithm of this paper.

Figure~\ref{fig:sbm}b shows a different test of the method.  Again we use
networks generated from a stochastic block model with two groups and
calculate the fraction of correctly classified nodes.  Now, however, we
vary the amount of noise introduced into the network to test the
algorithm's ability to recover structure in data of varying quality.  The
parameters of the underlying network are held constant, as is the
parameter~$c$ that controls the sparsity of the data matrix~$\mat{Q}$.
This leaves only one degree of freedom, which we take to be the
parameter~$b_1$ of the noise process (see Eq.~\eqref{eq:beta1}).

A network with little noise in the data is one in which true edges in the
underlying network are represented by probabilities~$Q_{ij}$ close to~1, in
other words by a noise distribution~$\beta_1(Q)$ with most of its weight
close to~1.  Such distributions correspond to small values of the
parameter~$b_1$.  Noisier data are those in which the values of
the~$Q_{ij}$ are smaller, approaching the values for the non-edges, thereby
making it difficult to distinguish between edges and non-edges.  These
networks are generated by larger values of~$b_1$.  Figure~\ref{fig:sbm}b
shows the fraction of correctly classified nodes as a function of~$b_1$, so
the noise level is increasing, and the quality of the simulated data
decreasing, from left to right in the figure.

As we can see, the algorithm returns close to perfect results when $b_1$ is
small---meaning that the quality of the data is high and the algorithm
almost sees the true underlying structure of the network.  Performance
degrades as the noise level increases, although the algorithm continues to
do significantly better than chance even for high levels of noise,
indicating that there is still useful information to be extracted even from
rather poor data sets.

\subsection{Protein interaction network}
\label{sec:ppi}
As a real-world example of our methods we have applied them to
protein-protein interaction networks from the STRING
database~\cite{VonMering2005}.  This database contains protein interaction
information for 1133 species drawn from a large body of research literature
covering a range of different techniques, including direct interaction
experiments, genomic information, and cross-species comparisons.  The
resulting networks are of exactly the form considered in this paper.  For
each network there is assumed to be a true underlying network in which
every pair of proteins either interacts or doesn't, but, given the
uncertainty in the data on which they are based, STRING provides only
probabilistic estimates of the presence of each interaction.  Thus the data
we have for each species consists of a set of proteins---the nodes---plus a
likelihood of interaction for each protein pair.  A significant majority of
protein pairs in each of the networks are recorded as having zero
probability of interaction, so the network is sparse in the sense assumed
by our analysis.

We analyze the data using the degree-corrected version of our algorithm
described in Section~\ref{sec:dcsbm}, which is appropriate because the
networks in the STRING database, like most real-world networks, have broad
degree distributions.

\begin{figure*}
\centering
\subfigure[~Method of this paper]{\includegraphics[width=7cm]{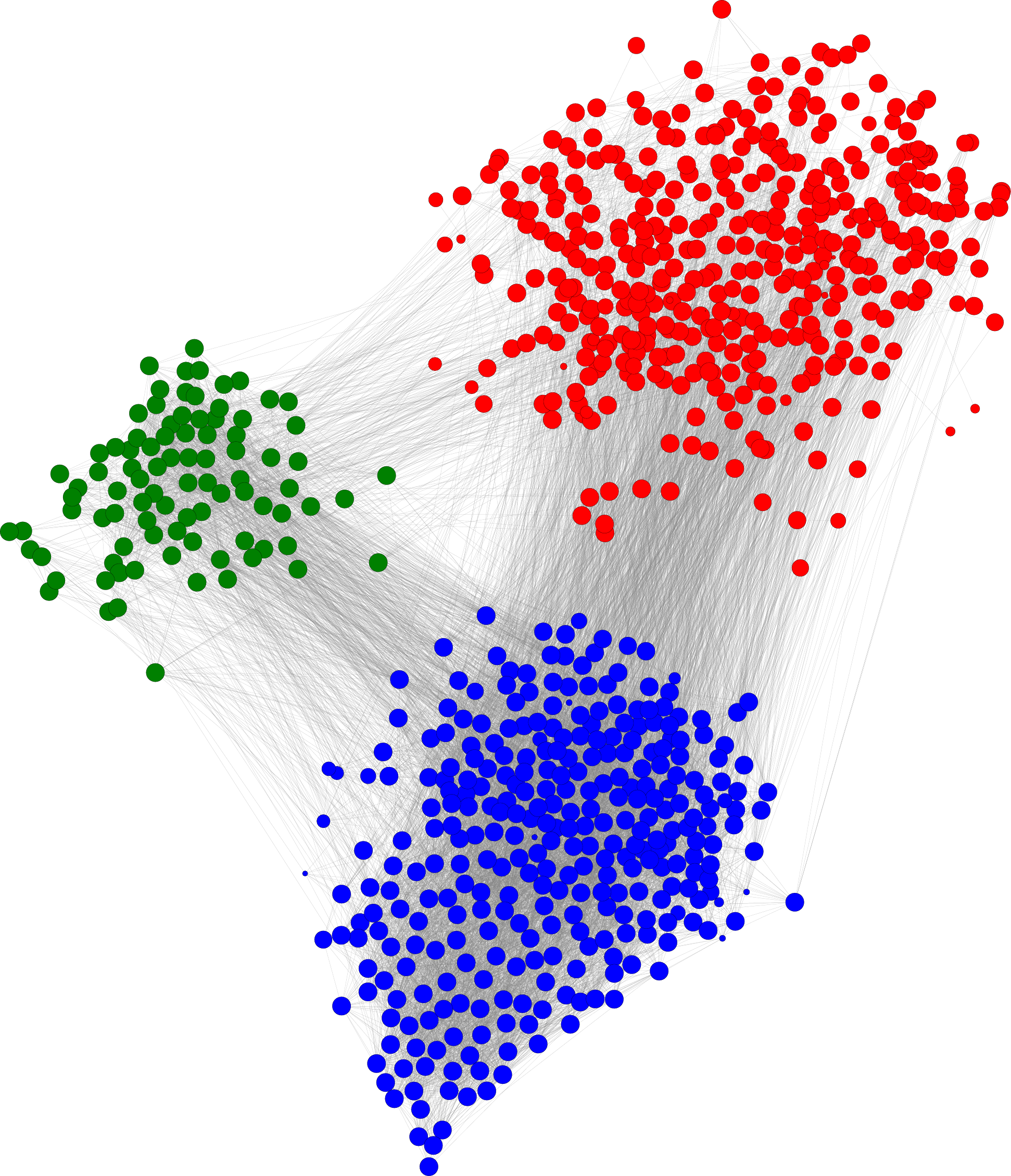}}
\hspace{4em}
\subfigure[~Thresholding method]{\includegraphics[width=7cm]{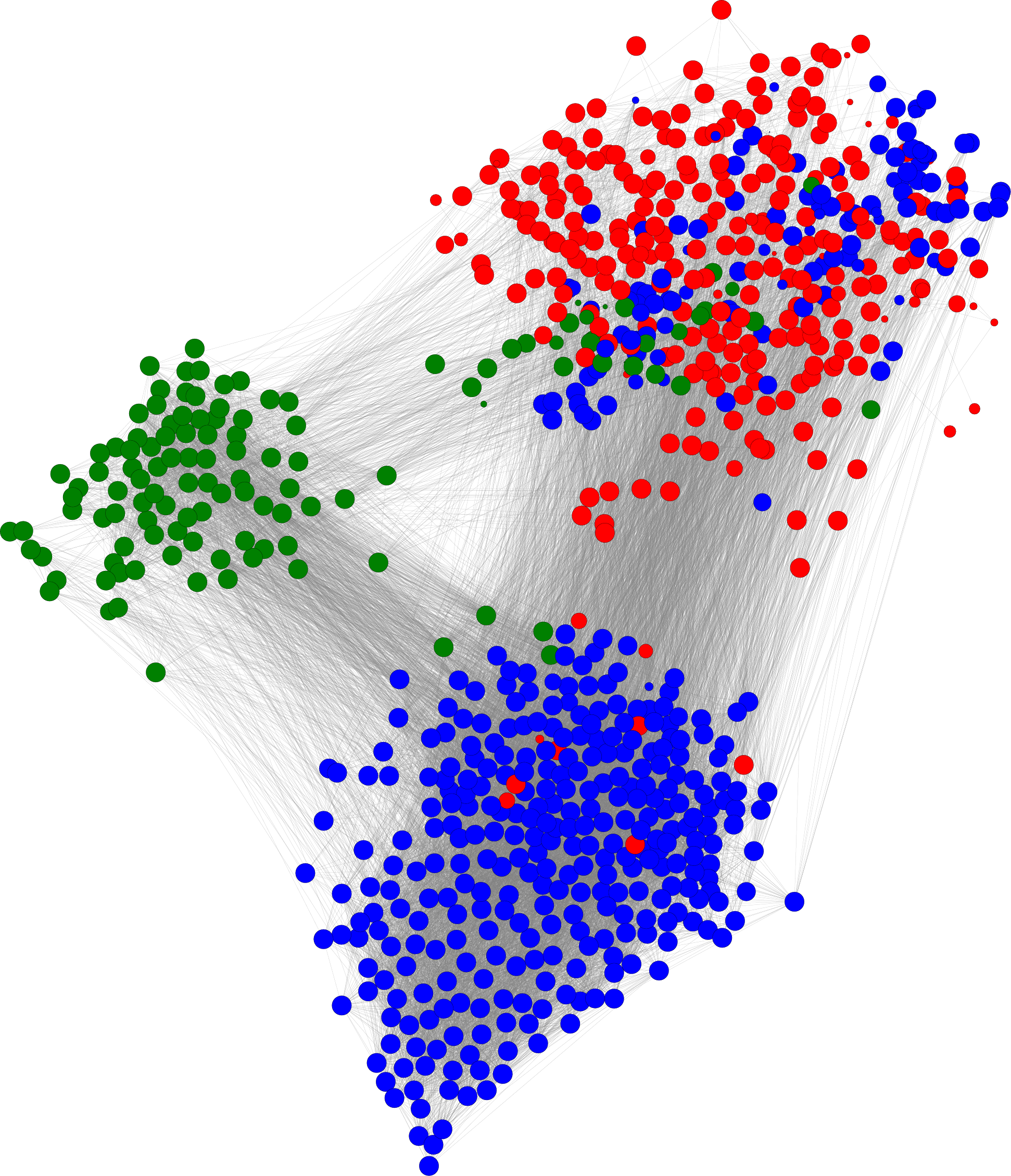}}
\caption{Communities found by (a)~the algorithm described in this paper and
  (b)~the thresholding algorithm, in a three-way split of the protein
  interaction network of the bacterium \textit{Borrelia hermsii}~HS1, taken
  from the STRING database.  Nodes are laid out according to the
  communities in (a) and the layout is the same in both panels.}
\label{fig:real}
\end{figure*}

Figure~\ref{fig:real}a shows the communities found in a three-way split of
the protein-protein interaction network of the bacterium \textit{Borrelia
  hermsii}~HS1.  Node colors denote the strongest community affiliation for
each node, as quantified by the one-node marginal probability~$q_r^i$, with
node size being proportional to the probability a node is in its most
likely community (so that larger nodes are more certain).  In practice,
most nodes belong wholly to just one community.

For comparison, we also show in Fig.~\ref{fig:real}b the communities found
in the same network by the naive thresholding algorithm discussed earlier
in which a node pair~$i,j$ is considered connected by an edge if and only
if the probability~$Q_{ij}$ exceeds a certain threshold, which here is set
at~0.25.  By contrast with the synthetic networks of the previous section,
we do not know the true underlying communities for this network and so
cannot calculate the fraction of correctly classified nodes, but it is
clear from the figures that the new technique gives significantly different
results from the thresholding method, particularly for the community that
appears in the upper right of the figure.

A closer examination of the data reveals a possible explanation.  The
communities at the left and bottom in both panels of Fig.~\ref{fig:real}
consist primarily of high-probability edges and are easily identified in
the data, so it is perhaps not surprising that both algorithms identify
these communities readily and are largely in agreement.  However, the third
community, in the upper right of the figure, consists largely of edges of
relatively low probability and the thresholding method has more difficulty
with this case because many edges fall below the threshold value and so are
lost, which may explain why the thresholding method divides the nodes of
this community among the three groups.

To give a simple picture, imagine a community whose nodes are connected by
very many internal edges, but all of those edges have low probability.
Because there are so many of them, the total expected number of true
internal edges in the underlying network---the number of node pairs times
the average probability of connection---could be quite high, high enough to
create a cohesive network community.  Our algorithm, which takes edge
probabilities into account, will allow for this.  The thresholding
algorithm on the other hand can fail because the edges all have low
probability, below the threshold used by the algorithm, and hence are
discarded.  The result is that the thresholding algorithm sees no edges at
all and hence no community.  The fundamental problem is that thresholding
is just too crude a tool to see subtle patterns in noisy data.

\section{Edge recovery}
A secondary goal in our analysis of uncertain networks is to deduce the
structure of the (unobserved) underlying network from the uncertain data.
That is, given the matrix~$\mat{Q}$ of edge probabilities, can we make an
informed guess about the adjacency matrix~$\mat{A}$?  We call this the
\defn{edge recovery} problem.  It is related to, but distinct from, the
well studied \defn{link prediction} problem~\cite{LK07}, in which one is
given a binary network of edges and non-edges but some of the data may be
erroneous and the problem is to guess which ones.  In the problem we
consider, by contrast, the data given are assumed to be correct, but they
are incomplete in the sense of being only the probabilities of the edges,
rather the edges themselves.

The simplest approach in the present case is simply to use the edge
probabilities~$Q_{ij}$ themselves to predict the edges---those node
pairs~$i,j$ with the highest probabilities are assumed most likely to be
connected by edges.  But if we know, or believe, that our network contains
community structure, then we can do a better job.  If we know where the
communities in the network lie, at least approximately, then given two
pairs of nodes with similar values of~$Q_{ij}$, the pair that are in the
same community should be more likely to be connected by an edge than the
pair that are not (assuming ``assortative'' mixing in which edge
probabilities are higher inside communities).

It turns out that our EM algorithm gives us precisely the information we
need to perform edge recovery.  The (posterior) probability of having an
edge between any pair of nodes~$i,j$ can be written as
\begin{align}
& P(A_{ij}=1) \nonumber\\
  &\qquad = \sum_{rs} P(A_{ij}=1 | g_i=r,g_j=s) P(g_i=r,g_j=s) \nonumber\\
  &\qquad = \sum_{rs} t_{rs}^{ij} q_{rs}^{ij},
\label{eq:edgerecovery}
\end{align}
where the data~$\mat{Q}$ and the
parameters~$\boldsymbol{\gamma},\boldsymbol{\omega}$ are assumed given in
each probability and we have made use of Eq.~\eqref{eq:tinterp} and the
definition of~$q_{rs}^{ij}$.  Both $t_{rs}^{ij}$ and $q_{rs}^{ij}$ are
calculated in the course of running the EM algorithm, so we already have
these quantities available to us and calculating $P(A_{ij}=1)$ is a small
extra step.

\begin{figure}
\centering
\includegraphics[width=6.5cm]{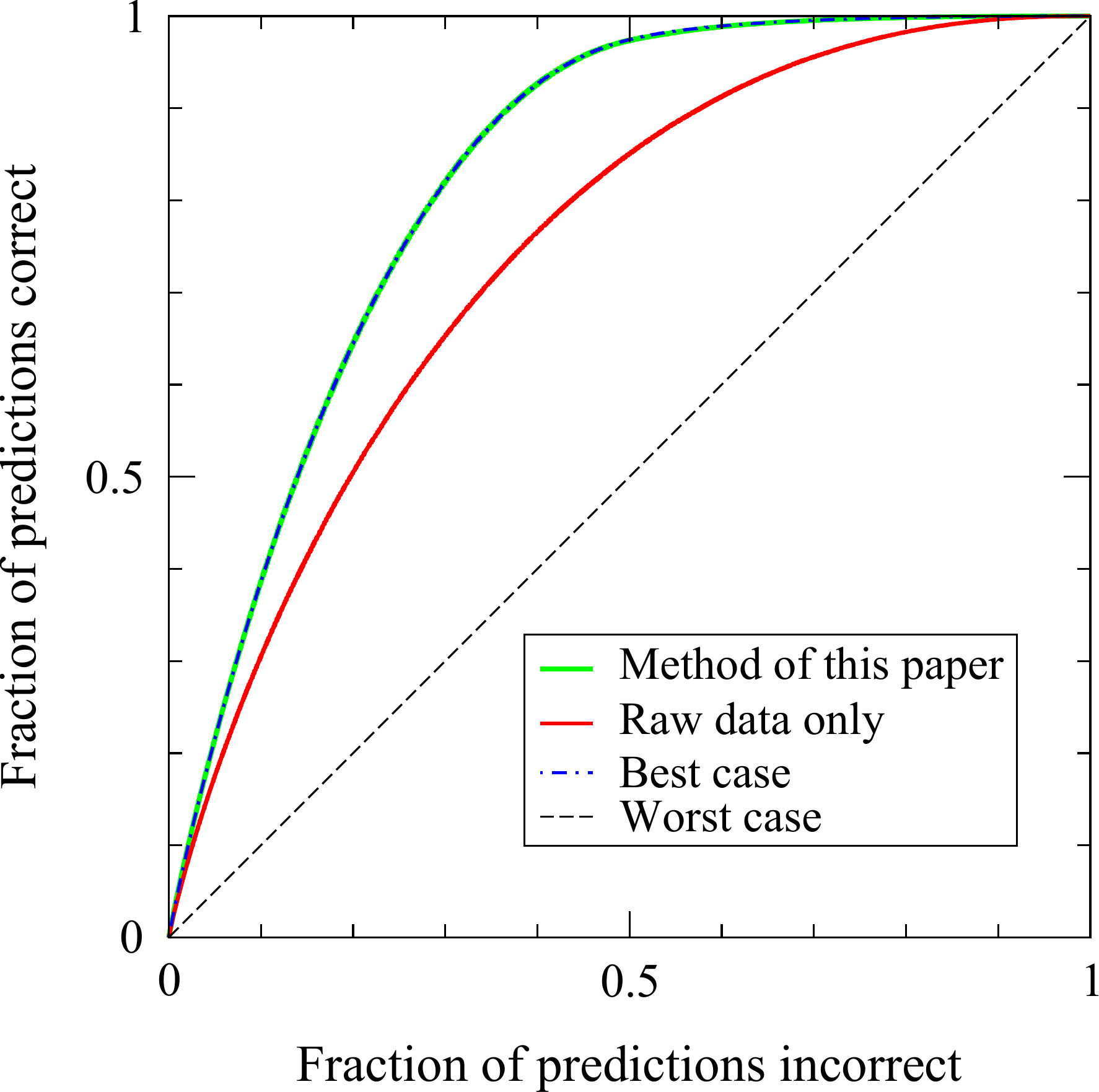}
\caption{Receiver operating characteristic (ROC) curves for the edge
  recovery problem on a synthetic network generated using a two-group
  stochastic block model with $n=4000$ nodes,
  $\omega_{11}=\omega_{22}=0.05$, $\omega_{12}=\omega_{21}=0.001$, and
  noise parameters~$b_1 = 4$ and $c=1/4n$.  The three curves show the
  performance of the algorithm of this paper, the naive algorithm based on
  the raw probabilities~$Q_{ij}$ alone, and a hypothetical ``ideal''
  algorithm that knows the values of the parameters used to generate the
  model (so that one does not have to run the EM algorithm at all).  The
  diagonal dashed line represents is curve generated by an algorithm that
  does no better than chance.}
\label{fig:edge_predict}
\end{figure}

Figure~\ref{fig:edge_predict} shows a test of the accuracy of our edge
predictions using synthetic test networks once again.  In these tests we
generate networks with community structure using the standard stochastic
block model, as previously, then run the network through the EM algorithm
and calculate the posterior edge probabilities of
Eq.~\eqref{eq:edgerecovery} above.  We compare the results against
competing predictions based on the prior edge probabilities~$Q_{ij}$ alone.

The figure shows \defn{receiver operating characteristic} (ROC) curves of
the results.  To construct an ROC curve, one asks how many edges we would
get right, and how many wrong, if we were to simply predict that the
fraction~$x$ of node pairs with the highest probabilities of connection are
in fact connected by edges.  The ROC curve is the plot of the fraction of
such predictions that turn out right (true positives) against the fraction
wrong (false positives) for values of $x$ from zero to one.  By definition
the curve always lies on or above the 45-degree line and the higher the
curve the better the results, since a higher curve implies more true
positives and fewer false ones.

Figure~\ref{fig:edge_predict} shows the ROC curves both for our method and
for the naive method based on the raw probabilities~$Q_{ij}$ alone and we
can see that, for the particular networks studied here, the additional
information revealed by fitting the block model results in a substantial
improvement in our ability to identify the edges of the network correctly.
One common way to summarize the information contained in an ROC curve is to
calculate the area under the curve, where an area of $0.5$ corresponds to
the poorest possible results---no better than a random guess---and an area
of $1$ corresponds to perfect edge recovery.  For the example shown in
Fig.~\ref{fig:edge_predict}, the area under the curve for our algorithm is
$0.89$ while that for the naive algorithm is significantly lower at~$0.80$.

Also shown in the figure is a third curve representing performance on the
edge recovery task if we assume we know the exact parameters of the
stochastic block model that were used to generate the network, i.e.,~that
we don't need to run the EM algorithm to learn the parameter values.  This
is an unrealistic situation---we very rarely know such parameters in the
real world---but it represents the best possible prediction we could hope
to make under any circumstances.  And, as the figure shows, this best
possible performance is in this case indistinguishable from the performance
of our EM algorithm, indicating that the EM algorithm is performing the
edge recovery task essentially optimally in this case.

\section{Conclusions}
In this paper we have described methods for the analysis of networks
represented by uncertain measurements of their edges.  In particular we
have described a method for performing the common task of community
detection on such networks by fitting a generative network model to the
data using a combination of an expectation--maximization (EM) algorithm and
belief propagation.  We have also shown how the resulting fit can be used
to reconstruct the true underlying network by making predictions of which
nodes are connected by edges.  Using controlled tests on computer-generated
benchmark networks, we have shown that our methods give better results than
previously used techniques that rely on simple thresholding of
probabilities to turn indefinite networks into definite ones.  And we have
given an example application of our methods to a bacterial protein
interaction network taken from the STRING database.

The methods described in this paper could be extended to the detection of
other types of structure in networks.  If one can define a generative model
for a structure of interest then the developments of
Section~\ref{sec:methods} can be applied, simply replacing the
likelihood~$P(\mat{A},\vec{g}|\boldsymbol{\gamma},\boldsymbol{\omega})$ in
Eq.~\eqref{eq:pqggw} with the appropriate probability of generation.
Generative models have been recently proposed for hierarchical structure in
networks~\cite{CMN08}, overlapping communities~\cite{ABFX08}, ranking or
stratified structure~\cite{BN13}, and others.  In principle, our methods
could be extended to any of these structure types in uncertain networks.

\begin{acknowledgments}
  The authors thank Michael Wellman and Erik Brinkman for useful comments
  and suggestions.  This research was funded in part by the US National
  Science Foundation under grants DMS--1107796 and DMS--1407207.
\end{acknowledgments}

\end{document}